\documentclass{aa}
\usepackage{graphics}
\begin{document}

\def\frac{$''$\hspace*{-.1cm}}
\def\deg{$^{\circ}$}
\def\min{$'$}
\def\deg{$^{\circ}$\hspace*{-.1cm}}
\def\min{$'$\hspace*{-.1cm}}
\def\h2{H\,{\sc ii}}
\def\hi{H\,{\sc i}}
\def\hb{H$\beta$}
\def\ha{H$\alpha$}
\def\oiii{[O\,{\sc iii}]}
\def\sm{$M_{\odot}$}
\def\ab{$\sim$}
\def\x{$\times$}
\def\sec{s$^{-1}$}
\def\lh5{LH\,5}

\title{
{\it HST} study of the LMC compact star forming region N83B\thanks{Based 
   on observations with 
   the NASA/ESA Hubble Space Telescope obtained at the Space Telescope 
   Science Institute, which is operated by the Association of Universities 
   for Research in Astronomy, Inc., under NASA contract NAS\,5-26555.}}

\offprints{M. Heydari-Malayeri, heydari@obspm.fr}

\date{Received 8 January 2001 / Accepted 21 February 2001}

\titlerunning{LMC N83B}
\authorrunning{Heydari-Malayeri et al.}

\author{M. Heydari-Malayeri\inst{1} 
	\and 
  	V. Charmandaris \inst{2}
	\and 
  	L. Deharveng \inst{3}
	\and
	M.\,R. Rosa \inst{4,}\,\thanks{
    Affiliated to the Astrophysics Division, Space Science Department of
    the European Space Agency.}
	\and   
	D. Schaerer \inst{5} 
	\and  
    	H.  Zinnecker \inst{6}  }

\institute{{\sc demirm}, Observatoire de Paris, 61 Avenue de l'Observatoire, 
F-75014 Paris, France 
\and 
Cornell University, Astronomy Department, 106 Space Sciences Bldg.,
Ithaca, NY 14853, USA
\and
Observatoire de Marseille, 2 Place Le Verrier, F-13248 Marseille Cedex
4, France
\and
Space Telescope European Coordinating Facility, European Southern
Observatory, Karl-Schwarzschild-Strasse-2, D-85748 Garching bei
M\"unchen, Germany
\and 
Observatoire Midi-Pyr\'en\'ees, 14, Avenue E. Belin, F-31400 Toulouse,
France
\and 
Astrophysikalisches Institut Potsdam, An der Sternwarte 16, D-14482
Potsdam, Germany}

\abstract{
High resolution imaging with the {\it Hubble Space Telescope} uncovers
the so far hidden stellar content and the nebular features of the high
excitation compact \h2\, region N83B in the Large Magellanic Cloud
(LMC). We discover that the \h2\, region is powered by the
most recent massive starburst in the OB association \lh5 and
the burst has created about 20 blue stars spread over \ab\,30\frac\,
on the sky (7.5 pc). Globally N83B displays a turbulent
environment typical of newborn massive star formation sites. It
contains an impressive ridge, likely created by a shock and a cavity
with an estimated age of only \ab\,30,000 yr, sculpted in the ionized
gas by the powerful winds of massive stars. The observations bring to
light two compact \h2\, blobs, N83B-1 and N83B-2, and a small
arc-nebula, N83B-3 lying inside the larger \h2\, region. N83B-1, only
\ab\,2\frac.8 (0.7 pc) across, is the brightest and most
excited part of N83B. It harbors the presumably hottest star of the
burst and is also strongly affected by dust with an extinction of
$A_{V}=2.5$ mag. The second blob, N83B-2, is even more compact, with a
size of only \ab\,1\frac\, (0.3 pc). All three features are formed in
the border zone between the molecular cloud and the ionized gas
possibly in a sequential process triggered by the ionization front of
an older \h2\, region.  Our {\it HST} imaging presents an interesting
and rare opportunity to observe details in the morphology of the star
formation in very small spatial scales in the LMC which are in
agreement with the concept of the fractal structure of molecular star
forming clouds. A scenario which supports hierarchical massive star
formation in the LMC OB association \lh5\, is presented.
}

\maketitle

\section{Introduction}

The giant \h2\, complex N83 (see Fig.\,\ref{lh5} and
Table\,\ref{names}), which lies in the western part of the Large
Magellanic Cloud (LMC) symmetrically opposed to the famous 30 Doradus,
is a rather scarcely studied star formation site. Henize (\cite{hen})
identified four ionized gas regions lying in that direction (N83\,\,A,
B, C, D). Davies et al. (\cite{dav}) catalogued the whole gas complex
as DEM\,22 and found two more components (DEM\,22a and DEM\,22d).
Radio continuum observations towards N83 (McGee \& Newton
\cite{mc}) revealed the presence of a bright source (MC\,16) coinciding
with N83A (Caplan \& Deharveng \cite{cd85}).  Lucke and Hodge
({\cite{lh}) detected an OB association (\lh5) spanning over 16 square
minutes (3600 square pc) around N83. It contains 26 blue stars (Lucke
\cite{lucke}), the brightest of which is Sk-69$^{\circ}$30 (Sanduleak
\cite{sk}), a G5 Ia with a magnitude of $V=10.18$ (Massey et
al. \cite{massey}; Schmidt-Kaler et al. \cite{schmidt}).  There is
also a very faint early type Wolf-Rayet star of the nitrogen sequence,
BAT99-5 (Breysacher et al. \cite{bat} and references
therein). \\

\begin{table*}[t]	
\caption[]{Cross identifications for the LMC N83 objects shown in Fig.\,\ref{lh5}}	
\label{names}	
\begin{flushleft}
\begin{tabular}{llll} 	
\hline	
R.A.(J2000) &     Dec.(J2000) &      Names  & References \\     
\hline
04 54 03  &    --69 12 04   &    N83A, NGC\,1743, SL\,87,ESO\,56EN21 & Bica et al. \cite{bica} \\
04 54 23  &    --69 11 05   &    N83B, NGC\,1748, IC2\,114,LMC-DEM\,22c, ESO\,56EN24 & '' \\
04 54 01  &    --69 09 25   &    N83C & '' \\
04 53 56  &    --69 10 25   &    DEM22a, NGC\,1737, ESO\,56EN20 & '' \\
04 54 27  &    --69 09 47   &    DEM22d & '' \\
04 54 06.26 &  --69 11 59.6 &    Sk-69$^{\circ}$25   & Simbad \\
04 54 14.26 &  --69 12 36.4 &    Sk-69$^{\circ}$30, HD\,268757, RMC\,59  & '' \\
04 54 28.05 &  --69 12 50.1 &    BAT99-5, Brey-4 & Breysacher et al. \cite{bat} \\
\hline
\end{tabular}
\end{flushleft}
\end{table*}

IRAS detected two infrared sources towards N83A and N83B (Israel \&
Koornneef \cite{ik91}), and a study of 69
\h2\, regions by Ye (\cite{ye}), using radio continuum observations
obtained at the Molonglo Observatory Synthesis Telescope 
(MOST) and H$\alpha$ fluxes, showed that  N83A ranks
among the four most reddened regions in the LMC where ionized
gas is associated with dust. Although N83B was not
included in the survey, these findings argued in favor of an 
important star formation region. \\

High-resolution neutral atomic hydrogen observations have recently
shown a supergiant \hi\, shell in that part of the LMC (Kim et
al. \cite{kim}). Since the shells are believed to be created by the
combined effects of the radiation pressure and the stellar wind of
massive stars on the cool interstellar gas, these observations
supported the presence of very hot stars towards N83. \\

Using the 2.2m and 3.6m telescopes of the European Southern
Observatory (ESO), Heydari-Malayeri et al. (\cite{paperi}, hereafter
Paper I) discovered a compact \h2\, region of \ab\,5\frac\, in size
towards N83B.  This object, which they named N83B-1, turned out to be
a member of the so-called high-excitation blobs (HEBs) in the
Magellanic Clouds.  In contrast to the typical \h2 regions of the
Magellanic Clouds, which are extended structures spanning several
minutes of arc on the sky and powered by a large number of hot stars,
HEBs are very dense small regions usually 5\frac\, to 10\frac\, in
diameter.  At the distance of the Magellanic Clouds this corresponds
to sizes of more than 50 pc for normal \h2\, regions and 1 to 3 pc for
the blobs. HEBs are in fact associated with young massive stars just
leaving their pre-natal molecular cloud (see Heydari-Malayeri et
al. \cite{hey1999c} for references). Paper I investigated several
physical characteristics of N83B-1, such as the emission spectrum,
excitation, extinction, gas density, chemical composition, abundances,
etc. Comparison of {\it JHK} photometry and IRAS spectra indicated the
presence of a compact infrared object with apparent dimensions of the
order of 10\frac\, towards N83B (Israel \& Koornneef \cite{ik91}). \\

In this paper we use observations obtained with the {\it Hubble Space
Telescope} to study the compact \h2\, region N83B.  The higher
resolution of {\it HST} is essential in order to reveal the various
emission and dust features of the nebula. Of particular interest is also
to study the stellar content of the \h2\, region and our ability to unveil
and identify the exciting stars, which up to now have  remained
unknown. \\

\begin{figure*}
\begin{center}
\resizebox{17cm}{!}{\includegraphics{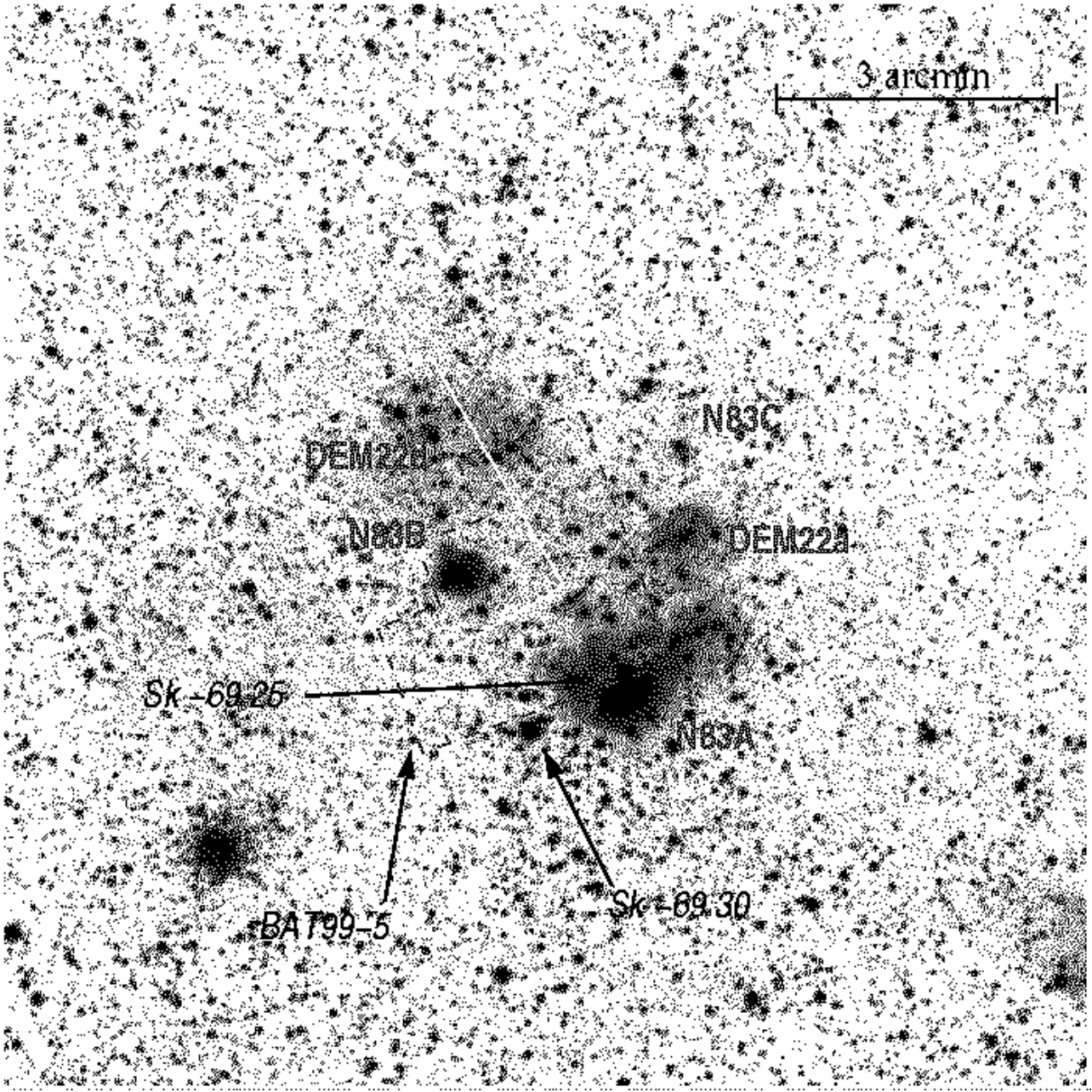}}
\caption{The LMC OB association \lh5\, with its central giant 
\h2\, complex N83, as seen on the ESO $R$ survey (Simbad server). 
See Table\,\ref{names} for cross identifications.  The ionized
components and three stars of the association are labeled.  The
marked stars are Sk-69$^{\circ}$30, the brightest star of the
association, Sk-69$^{\circ}$25, the main exciting star of N83A, and
BAT99-5, a Wolf-Rayet star (see the text).  The WFPC2 footprints are
also indicated. The solid line shows the field of view during the {\it
HST} imaging with the nebular filters (21 May 2000) and the dashed
line is the one for the stellar filters (7 February 2000). The field
center is at: $\alpha$\,=\, 04:54:14.48, $\delta$\,=\,--69:10:33.7
(J2000). }
\end{center}
\label{lh5}
\end{figure*}

\begin{figure*}
\begin{center}
\resizebox{11cm}{!}{\includegraphics{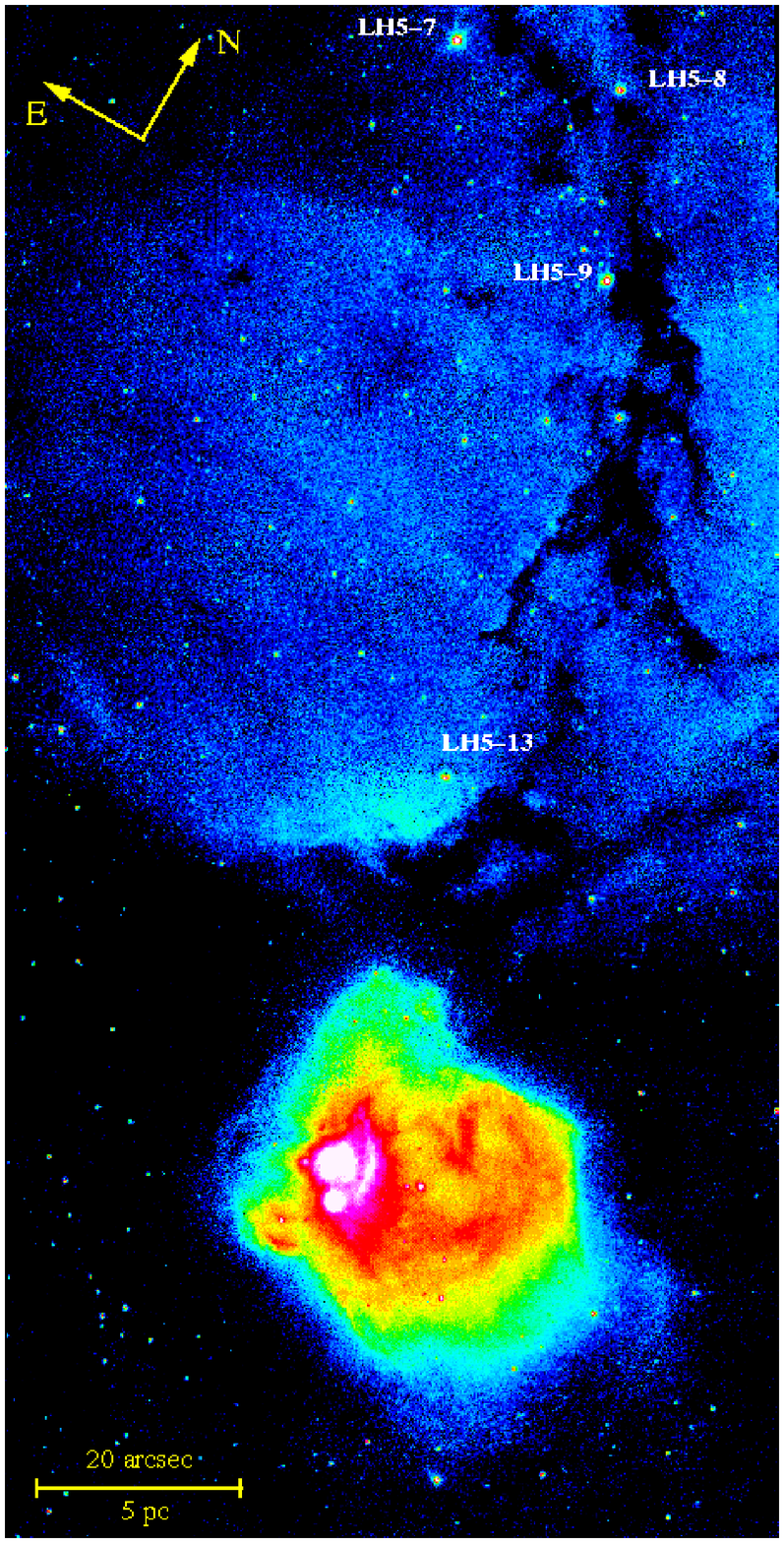}}
\caption{A large view of the LMC compact nebula N83B and its
neighboring field imaged with the {\it HST} camera WFPC2 in the
hydrogen recombination line \ha\, (filter F656N).  See
Fig.\,\ref{true} for details of the compact \h2\, region. The field
size is \ab\,66\,\frac\,\x\,133\frac\, (\ab\,17\,pc\,\x\,33\,pc) and
comprises the images by the CCDs WF2 and WF3.  The four bright,
labelled stars are known members of the OB association \lh5. Downward
from top, their $V$ magnitudes are 13.82, 14.99, 14.16, and 15.16
respectively.}
\label{habig}
\end{center}
\end{figure*}

\section{Observations and data reduction}

The observations of N83B were performed with the Wide Field Planetary
Camera 2 (WFPC2) on board of the {\it HST} using several broad- and
narrow-band filters. The images taken with the broad-band filters
(F300W, F467M, F410M, and F547M) were obtained on February 7, 2000 and
had as a goal to reveal the details of the stellar content of N83B
which was centered on the Planetary Camera (PC). The narrow-band
filter images (F487N, F503N and F656N) were obtained on May 21,
2000. In that case the target was centered on the WF2 which has larger
pixels and lower noise than the PC CCD and is better suited for
detecting faint nebular emission. Exposures were taken at 4 different
pointings offset by 0\frac.8 and the exposure times ranged from 10 to
300 sec (see Table\,\ref{obs} for details). \\

\begin{table}[!h]  
\caption[ ]{N83B observations ({\it HST} GO-8247)} 
\label{obs} 
\begin{flushleft}  
\begin{tabular}{lcr}  
\hline 
{\it HST} filter              & Wavelength    & Exposure time\\ 
                        & $\lambda$(\AA)        & (sec)\\
\hline 
F300W (wide-U)                  & 2911          & 4\,\x\,14\,=\,56\\
F410M (Str\"{o}mgren $v$)       & 4090          & 4\,\x\,50\,=\,200\\
F467M (Str\"{o}mgren $b$)       & 4669          & 4\,\x\,35\,=\,140\\
F547M (Str\"{o}mgren $y$)       & 5479          & 4\,\x\,10\,=\,40\\
F487N (H$\beta$)                & 4866          & 4\,\x\,260\,=\,1040\\
F502N ([OIII])                  & 5013          & 4\,\x\,300\,=\,1200\\
F656N  (H$\alpha$)              & 6563          & 4\,\x\,260\,=\,1040\\
\hline    
\end{tabular} 
\end{flushleft}   
\end{table}

The data were processed through the standard {\it HST} pipeline
calibration.  Multiple images were co-added using the {\sc stsdas}
task {\it imcombine}, while cosmic rays were detected and removed with
the {\sc stsdas} task {\it crrej}.  Normalized images were then
created using the total exposure times for each filter.  To extract
the positions of the stars, the routine {\it daofind} was applied to
the images by setting the detection threshold to 5$\sigma$
above the local background level.  The photometry was performed
setting a circular aperture of 3--4 pixels in radius in the {\it
daophot} package in {\sc stsdas}. \\

\begin{figure*}[!ht]
\begin{center}
\resizebox{15cm}{!}{\includegraphics{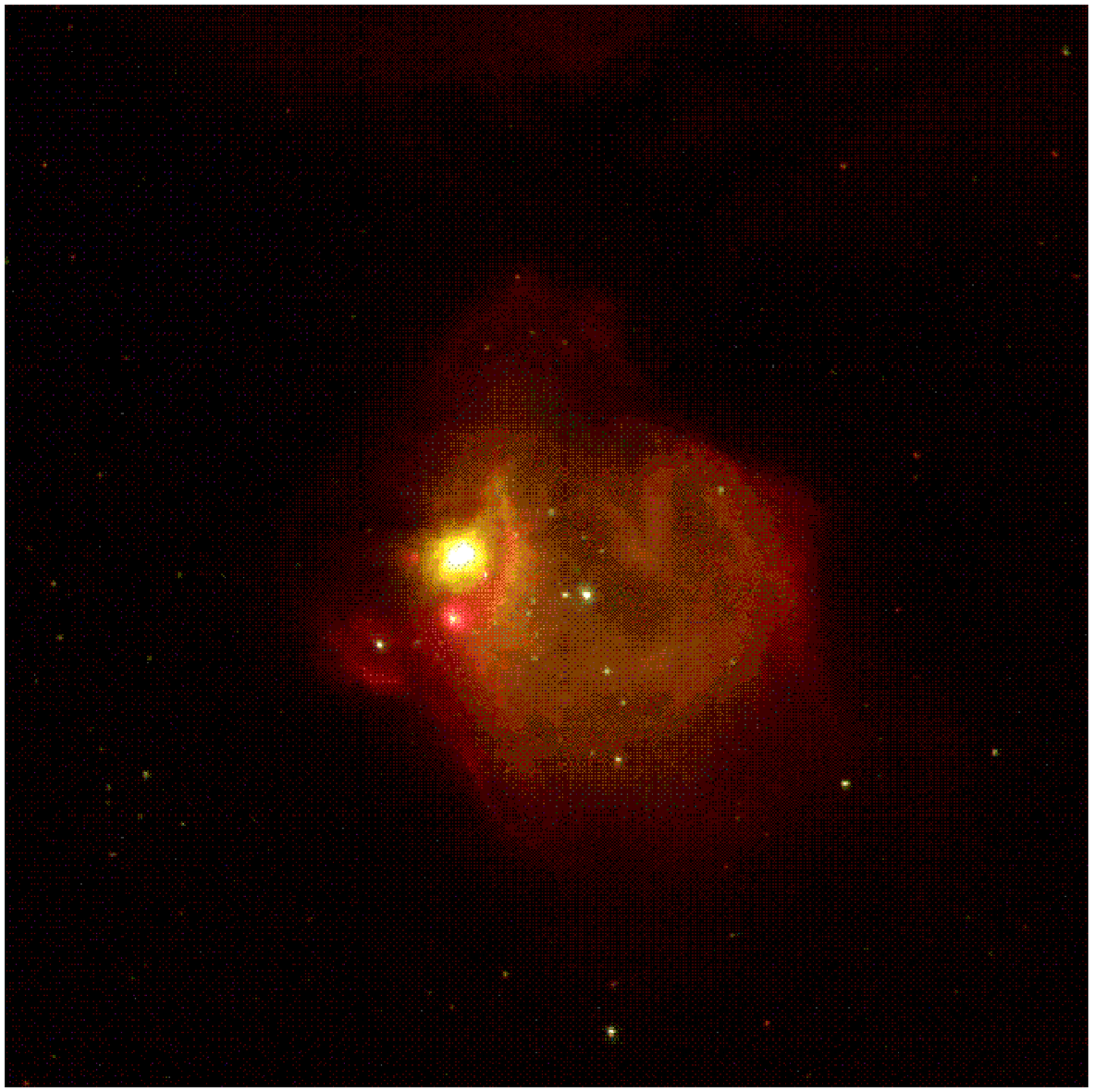}}
\caption{A ``true color'' composite image of the LMC compact \h2\,
region N83B as seen by {\it HST}/WFPC2, based on images taken with
filters \ha\, (red), \oiii\, (green), and \hb\, (blue).  The
brightest blob is the compact \h2\, region N83B-1
(\ab\,2\frac.8 across, 0.7 pc, surrounding star \#29 ) and the fainter
one below it (\ab\,1\frac\, in size surrounding star \#28) is N83B-2.
The small arc-nebula further south, centered on a relatively bright
star (\#34, see also Fig.\,\ref{chart}), is N83B-3.  Note
also the outstanding shock ridge and the cavity to the west of N83B-1,
sculpted in the ionized gas of N83B by the stellar wind of the central
bright star (\#10), with its shell structure and curl.  The field size
is \ab\,63\frac\,\x\,63\frac\, (\ab\,15 pc \x\,15 pc), and the
orientation the same as in Fig.\,\ref{habig}.}
\label{true}
\end{center}
\end{figure*}

A crucial point in our data reduction was the sky subtraction. For
most isolated stars the sky level was estimated and subtracted
automatically using an annulus of 6--8 pixel width around each star.
However this could not be done for several stars located in the
central region of N83B due to their crowding. In those cases we
carefully examined the PSF size of each individual star ({\sc
fwhm}\,\ab\,2 pixels, corresponding to 0\frac.09 on the sky) and did
an appropriate sky subtraction using the mean of several nearby
off-star positions.  To convert into a magnitude scale we used zero
points in the Vegamag system, that is the system where Vega is set to
zero mag in Cousin broad-band filters.  The magnitudes measured were
corrected for geometrical distortion, finite aperture size (Holtzman
et al. \cite{holtz}), and charge transfer efficiency as recommended by
the {\it HST} Data Handbook. Our broad-band images reveal 179
stars in the WFPC2 field of view. Thirty six (36) of those are
confined towards N83B-1 within the area covered the PC. Most of them
are also visible in the true-color image (Fig.\,\ref{true}), and can
be identified using the finder chart presented in
Fig.\,\ref{chart}. In Table\,\ref{phot} we summarize the photometry
for those stars around N83B-1 (the PC field of view) which are
brighter than 19th magnitude in the Str\"omgren $y$ filter, as we
cannot provide accurate colors for the fainter ones. The photometric
errors estimated by {\it daophot} are smaller than 0.01 mag for the
brighter (14--15 mag) stars, while they increase to $\sim$\,0.2 mag
for 19 mag stars. \\

We note that the filter F547M is wider than the standard Str\"omgren
$y$ filter. To evaluate the presence of any systematic effects in our
photometry and color magnitude diagrams due to this difference in the
filters, we used the {\sc stsdas} package {\it synphot}.  Using
synthetic spectra of hot stars, with spectral types similar to those
found in \h2\, regions, we estimated the difference due to the {\it
HST} band-passes to be less than 0.002 mag, which is well within the
photometric errors. \\

\begin{figure*}[!ht]
\begin{center}
\resizebox{14cm}{!}{\includegraphics{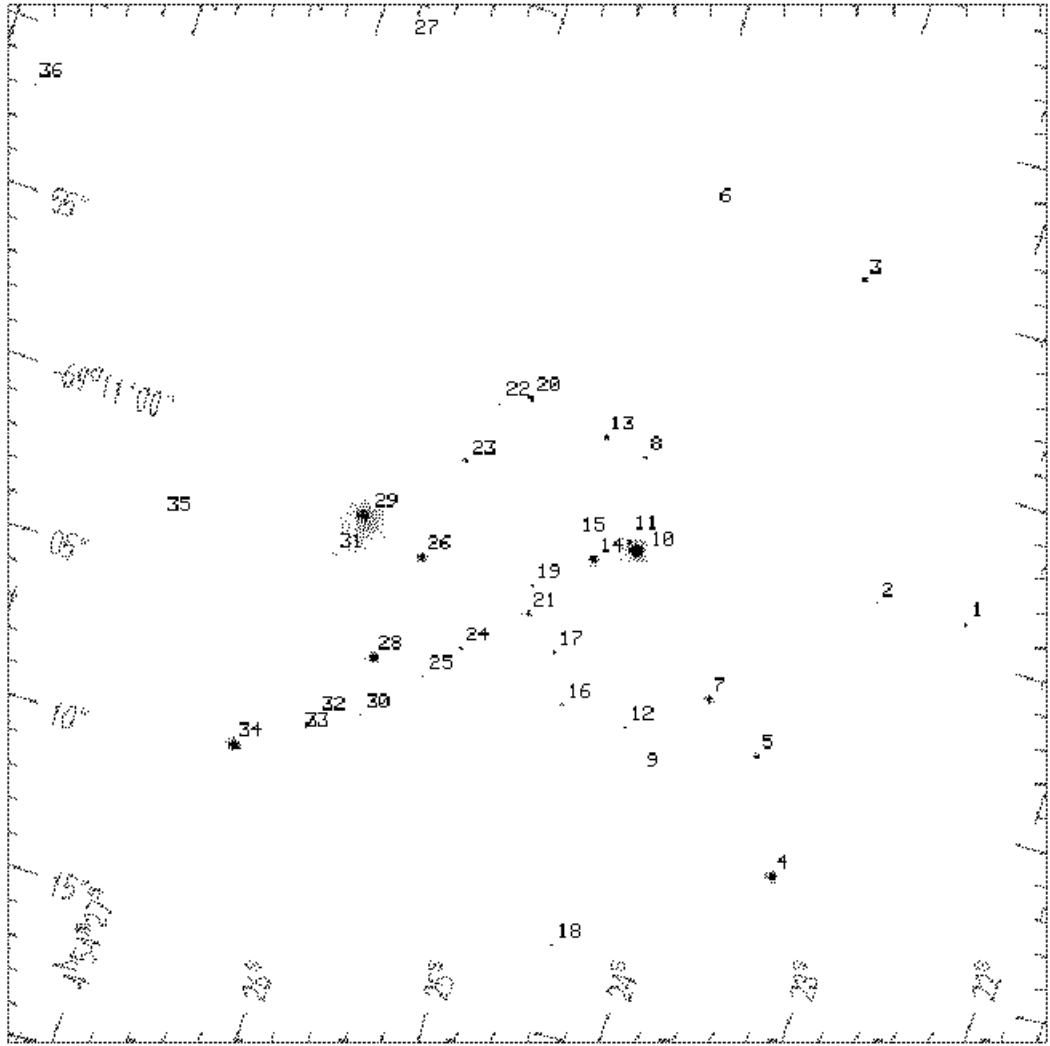}}
\caption{A WFPC2 image of the LMC N83B taken with the Strömgren $y$ filter 
(F547M) where the brighter stars are labeled.  The photometry of the
stellar subsample is presented in Table\,\ref{phot}. The
field size is \ab\,32\frac\,\x\,32\frac\, (\ab\, 8 pc \x\, 8 pc ),
and the coordinates are in J2000.}
\label{chart}
\end{center}
\end{figure*}

\section{Results}

\subsection{Morphology}

A large-scale image of the N83B field in H$\alpha$ emission obtained
with {\it HST} is shown in Fig.\,\ref{habig}, while a close-up view of the
\h2\, region is presented in Fig.\,\ref{true}.  The images for the
first time resolve N83B and reveal several interesting nebular and
stellar features. The \h2\, blob N83B-1, lies in the brightest part of
the nebula, and is nearly two times more compact than previously
believed (\ab\,2\frac.8 across, 0.7 pc).  The images also unveil
a previously unknown even smaller \h2\, region of
\ab\,1\frac\, in diameter lying less than 4\frac\, south-west of
N83B-1. Hereafter we will call it N83B-2. 
Another interesting ionized region,  N83B-3, situated some 6\frac.5 
(1.6 pc) south of N83B-1, is centered on star \#34 and presents a 
relatively bright circular filament or arc to the south.   \\

An outstanding feature, clearly visible in the above figures, is a
narrow ridge or partial shell of ionized gas centered on N83B-1 and
situated to its west at a projected distance of 2\frac.5 (0.6
pc). This ridge, most probably produced by powerful stellar winds and
shocks from the hot stars \#29 and \#10, indicates a very turbulent
environment typical of star forming regions.  All these emission
features around N83B-1 are found in the eastern part of a larger
diffuse nebula \ab\,30\frac\, (\ab\,7.5 pc) in diameter, N83B, which
displays a shell structure, the northern part of which is
abruptly curved to the east.  The low brightness at the center of
this shell nebula is most likely due to a cavity already created by
the brightest star of the region (that is star \#10, see Section 3.4). \\

Further north of N83B the much larger and even more diffuse
nebular region  DEM\,22d is visible (top of Fig.\,\ref{habig}). It is
crossed by a prominent dark absorption lane extending over more than a
minute of arc or 15 pc along the north-northeast direction.  This
structure is situated in front of the diffuse emission
nebula. Even though it is not clearly obvious in the printed
version of Fig.\,\ref{habig}, inspecting the dark lane in detail, it
appears that its borders are enhanced in brightness indicates that it
is dense and ionized from outside, probably by the radiation field of
the underlying nebula. This structure is somewhat reminiscent of ``the
gaseous pillars'' observed by {\it HST} in M16. \\

The filamentary and wind induced structures are best seen in
Fig.\,\ref{mask}, which presents an un-sharp masking image of N83B in
\ha\, from which large scale structures have been subtracted.  In
order to remove these brightness variations and enhance the high
spatial frequencies, a digital ``mask'' was created from the \ha\,
image.  First the \ha\, image was convolved by a 2\,\x\,2-pixel
Gaussian, and then the smoothed frame was subtracted from the original
\ha\, image.  Interestingly, the inspection of the orientation of
these arched filaments suggests the presence of several sources of
stellar winds, particularly stars \#29, \#34, and \#10 
which are among the brightest blue stars of the cluster.   \\

\begin{figure}[!h]
\resizebox{\hsize}{!}{\includegraphics{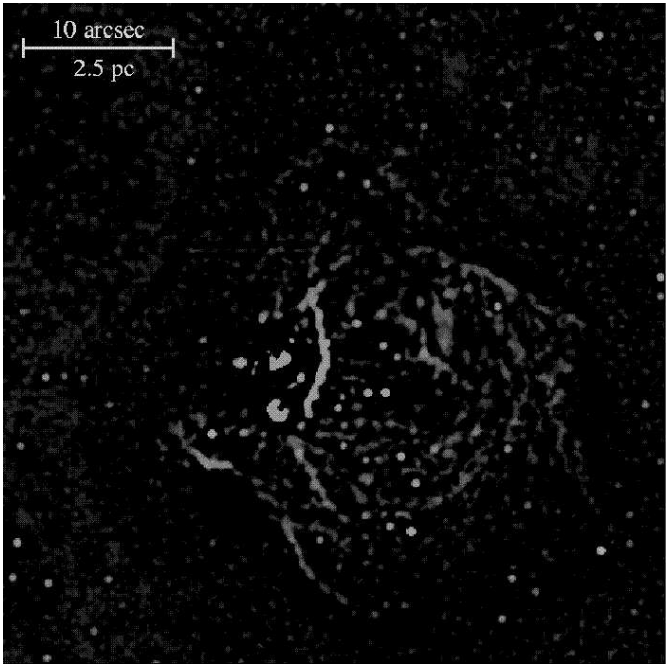}}
\caption{An un-sharp masking image of N83B obtained 
in \ha, which highlights the filamentary patterns of the nebula (see
the text).  The field size is \ab\,45\,\frac\,\x\,45\frac\,
(11\,pc\,\x\,11\,pc), and the orientation is as in Fig.\,4.}
\label{mask}
\end{figure}

\subsection{Extinction}

A map of the \ha/\hb\, Balmer decrement is presented in
Fig.\,\ref{rapports}a. It reveals a strong absorption zone coinciding
with the \h2\, blob N83B-1, where the ratio reaches values as high as
\ab\,7, while its mean value is 5.10\,$\pm0.30$.  Using the
interstellar reddening law, these ratios correspond to $A_{V}$\,=\,2.5
and 1.6 mag respectively. We note that star \#29, the second brightest
star in N83B (see Table\,\ref{phot}), is situated in the
direction of this high extinction region. Another peak with slightly
smaller reddening is centered on star \#28, located inside the compact
blob N83B-2, and reaches a mean \ha\,/\hb\, value of 4.5 corresponding
to $A_{V}$\,=\,1.3 mag. \\
 
Outside those areas the reddening is smaller and becomes almost
uniform towards the diffuse component with a mean value of \ab\,3.7
which corresponds to $A_{V}$\,=\,0.7 mag.  For the whole N83B the
global Balmer decrement is slightly larger, 3.85\,$\pm0.02$ or
$A_{V}$\,=\,0.8 mag. This value revises the 3.99 which had been
estimated in Paper I. The new \ha\,/\hb\, map was used to accurately
correct the \hb\, flux of the \h2\, region for interstellar
reddening. The correction was applied to the \hb\, image on a pixel by
pixel basis using straightforward mathematical operations. Our
interpretation of these results is presented in the following
section. \\

\begin{table*}[t]	
\caption[]{{\it HST} Photometry of the brightest stars towards N83B}	
\label{phot}	
\begin{flushleft}
\begin{tabular}{cccccccc} 	
\hline	
  Star & RA(J2000) & Dec(J2000) & F300W & F410M & F467M & F547M &
  Color\\ & & & Wide $U$ & Str\"omgren $v$ & Str\"omgren $b$ &
  Str\"omgren $y$ & $b-y$ \\
\hline	
   1  &  4:54:22.7  & -69:10:59.0 & 17.57 & 18.44 & 18.48 & 18.61 & -0.13 \\
   2  &  4:54:23.3  & -69:10:59.2  &      &       &       &       & \\
   3  &  4:54:23.9  & -69:10:49.9 & 17.07 & 18.17 & 18.19 & 18.17 & +0.02 \\
   4  &  4:54:23.4  & -69:11:08.2 & 14.74 & 16.17 & 16.35 & 16.40 & -0.05\\
   5  &  4:54:23.7  & -69:11:04.8 & 16.18 & 17.62 & 17.64 & 17.80 & -0.16\\
   6  &  4:54:24.8  & -69:10:49.2 &       &       &       &       & \\
   7  &  4:54:24.0  & -69:11:03.6 & 15.48 & 16.99 & 17.15 & 17.12 & +0.03 \\
   8  &  4:54:24.9  & -69:10:57.1 & 17.56 & 18.80 & 18.78 & 18.84 & -0.06\\
   9  &  4:54:24.3  & -69:11:06.5 &       &       &       &       &\\
  10  &  4:54:24.7  & -69:10:59.9 & 12.56 & 14.26 & 14.48 & 14.57 & -0.09 \\
  11  &  4:54:24.7  & -69:10:59.8 & 16.56 & 17.70 & 18.13 & 18.24 & -0.11\\
  12  &  4:54:24.4  & -69:11:05.2 &       &       &       &       &\\
  13  &  4:54:25.0  & -69:10:56.9 & 16.75 & 18.08 & 18.22 & 18.29 & -0.07 \\
  14  &  4:54:24.9  & -69:11:00.6 & 15.55 & 16.93 & 17.09 & 17.15 & -0.06\\
  15  &  4:54:25.0  & -69:11:00.2 &       &       &       &       &\\
  16  &  4:54:24.8  & -69:11:05.1 &       &       &       &       &\\
  17  &  4:54:24.9  & -69:11:03.7 &       &       &       &       &\\
  18  &  4:54:24.4  & -69:11:12.3 &       &       &       &       &\\
  19  &  4:54:25.2  & -69:11:01.9 & 17.41 & 18.77 & 18.80 & 18.76 & +0.04 \\
  20  &  4:54:25.5  & -69:10:56.5 & 16.29 & 17.65 & 17.87 & 17.78 & +0.09\\
  21  &  4:54:25.2  & -69:11:02.8 & 16.51 & 17.83 & 17.87 & 17.96 & -0.09\\
  22  &  4:54:25.7  & -69:10:56.9 &       &       &       &       &\\
  23  &  4:54:25.8  & -69:10:58.9 & 16.50 & 17.81 & 17.98 & 17.89 & +0.09\\
  24  &  4:54:25.5  & -69:11:04.4 & 17.90 & 19.00 & 18.93 & 19.00 & -0.07\\
  25  &  4:54:25.6  & -69:11:05.6 &       &       &       &       &\\
  26  &  4:54:25.8  & -69:11:02.1 & 15.28 & 16.63 & 16.94 & 16.97 & -0.03\\
  27  &  4:54:26.8  & -69:10:47.1 &       &       &       &       &\\
  28  &  4:54:25.9  & -69:11:05.5 & 14.96 & 16.51 & 16.72 & 16.69 & +0.03\\
  29  &  4:54:26.2  & -69:11:01.5 & 14.66 & 15.92 & 16.04 & 15.85 & +0.19\\
  30  &  4:54:25.9  & -69:11:07.3 &       &       &       &       &\\
  31  &  4:54:26.3  & -69:11:02.9 &       &       &       &       &\\
  32  &  4:54:26.2  & -69:11:08.1 &       &       &       &       &\\
  33  &  4:54:26.2  & -69:11:08.2 &       &       &       &       &\\
  34  &  4:54:26.5  & -69:11:09.4 & 14.11 & 15.52 & 15.80 & 15.86 & -0.06\\
  35  &  4:54:27.3  & -69:11:03.4 &       &       &       &       &\\
  36  &  4:54:28.8  & -69:10:51.9 &       &       &       &       &\\
\hline
\end{tabular}
\end{flushleft}
\end{table*}

\subsection{Nebular emission}

The [O\,{\sc iii}]$\lambda$5007/\hb\, intensity map
(Fig.\,\ref{rapports}b) reveals a remarkably extended high-excitation
zone where the O$^{++}$ ions occupy almost the same volume as H$^{+}$.
However, the most excited area belongs to the compact component
N83B-1, where the ratio reaches values as high as \ab\,6, while its
mean value is around 4.5. Interestingly, the highest excitation area
is at the same time the most reddened part of the nebula (Section
3.2), indicating the association of hot gas with local dust.  On the
other hand, the diffuse component is significantly less excited with a
mean value \ab\,2. 
The second blob, N83B-2, is of even  lower excitation, with 
\oiii/\hb\,\ab\,1, 
indicating that its associated exciting star, \#28, has a
later spectral type than that of star \#29, which excites N83-B1, in
agreement with the stars' respective positions in the color-magnitude
diagram (Fig.\,\ref{cm}). Surprisingly, \oiii/\hb\, is fairly low 
(\ab\,1.5) along the bright narrow ridge. One possibility is that this 
ridge is an ionization/shock front, seen nearly edge on. This structure 
is very reminiscent of the ``Bright Bar'' in the Orion Nebula, a region of
enhanced \ha\, surface brightness with a high [N\,{\sc ii}]/\ha\,
ratio (O'Dell 2001, and references therein). \\

The total \hb\, flux of N83B was derived using the method explained in
Section 3.2.  The corrected flux is
$F_{0}$(\hb)\,=\,1.76\,$\times$\,10$^{-11}$ erg cm$^{-2}$ s$^{-1}$
above 3$\sigma$ level without the stellar contribution and accurate to
3\,\%.  Assuming that the \h2\, region is ionization-bounded, the
corresponding Lyman continuum flux of the region is
$N_{L}$\,=\,1.37\,$\times$\,10$^{49}$ photons s$^{-1}$.  A single main
sequence star of type O6.5 or O7 or several massive stars of later
types can account for this ionizing UV flux (Vacca et
al. \cite{vacca}, Schaerer \& de Koter \cite{sch}).  However, these
are apparently underestimates since the \h2\, region is more likely to
be density-bounded, as the west side of it, i. e.  the diffuse shell
nebula, appears to be torn open towards the interstellar medium.  \\

\begin{figure*}
\resizebox{\hsize}{!}{\includegraphics{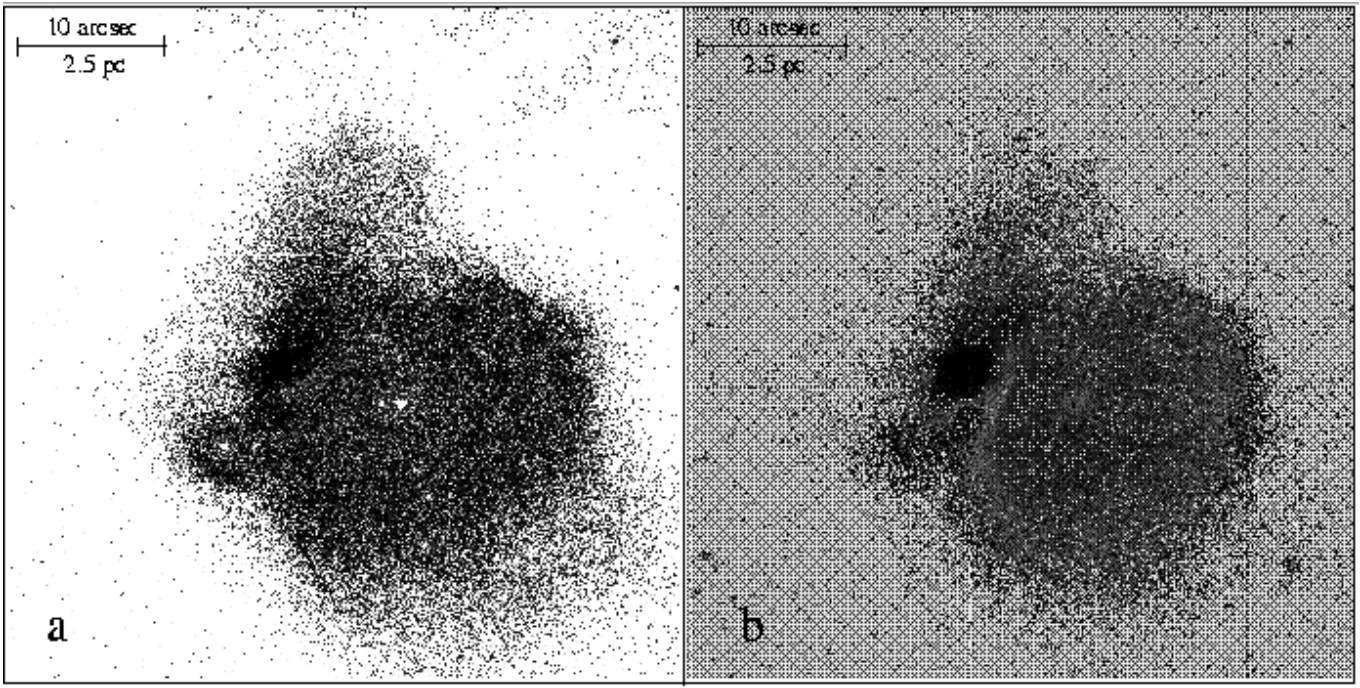}}
\caption{Line intensity ratios for the LMC compact nebula  
N83B. Darker colors correspond to higher ratio values.  The field of
view and orientation are identical to Fig.\,\ref{mask}. The white
spots are stars and can be identified using Fig.\,\ref{chart}.  {\bf
a)} Balmer decrement \ha/\hb. Its mean value over the diffuse
component is \ab\,3.7 ($A_{V}$\,=\,0.7 mag), while the ratio goes up
to \ab\,7 ($A_{V}$\,=\,2.5 mag) towards the compact blob N83B-1.  {\bf
b)} The [O\,{\sc iii}]$\lambda$5007/\hb\, ratio, where the highest
values, towards N83B-1, reach \ab\,6. }
\label{rapports}
\end{figure*}

\subsection{Stellar content}

The stars of the region imaged by HST have magnitudes ranging from
$y=14.57$ to \ab\,22.  Using a  magnitude cutoff of $y$\,=\,19, we construct
a color-magnitude (C-M) diagram of $y$ versus $b -y$  (Fig.\,\ref{cm})
in which we have also added all the background stars around N83B visible
in the whole field of WFPC2.  The C-M diagram displays
a rather well-defined main sequence for the N83B cluster in the
interval $14.56 \leq y \leq 19.00$ centered on colors $b - y = -0.04$,
or $v - b = -0.13$. The colors and absolute magnitudes (see below) are
typical of massive OB stars (Relyea \& Kurucz \cite{rk}, Conti et
al. \cite{conti}, Vacca et al. \cite{vacca}). There is no evidence for
stars evolved off the main sequence in N83B.  We note that star \#29
is offset red-wards ($b-y=+0.19$) in the C-M diagram. However, since
it lies in the most heavily obscured area of the N83B-1 blob, it is
quite probable that its color is reddened due to extinction by dust 
(see Section 3.2), and star \#29 may be in fact intrinsically blue
and one of the main exciting stars of the region. Similar reasoning
could explain the rather red colors of two more stars, \#23 which is
situated on the bright ridge and \#20 found just to the north side of
the ridge. \\

The brightest star observed is \#10 situated towards the center of the
diffuse shell nebula. It has a faint visual companion,
\#11, lying just 0\frac.3 to its east. The second brightest star is
\#94 which is a field  star apparently associated with the main
ionized gas component N83A; its J2000 coordinates are:
$\alpha$\,=\,4:54:15.6, $\delta$\,=\,--69:12:14.9.  Following in
brightness are stars \#29 and \#34 of N83B, the latter of which
ionizes the nebula N83B-3.  Another noteworthy star is \#28 which lies
inside the small nebular blob N83B-2. It should be a massive star
responsible for the ionization of the compact nebula. Stars \#34 and
\#28 are less reddened than star \#29 in agreement with the fact that
the extinction of their associated nebulae, estimated from the Balmer
decrement, is smaller than that of N83B-1. \\

One could try to estimate the luminosities of the brightest stars of
the cluster (\#10 and \#29), although in the absence of spectroscopic
data this would not be accurate. In the case of star \#10,
using an extinction of $A_{V}$\,=\,0.7 mag corresponding to
the mean value for the associated nebula (Section 3.2), and a distance
modulus $m$\,--\,$M$\,=\,18.5 (e.g.  Kov\'acs \cite{ko} and references
therein), we find a visual absolute magnitude $M_{V}=-4.63$.  If the
star is on the main sequence, according to the calibration of Vacca et
al. (\cite{vacca}) for Galactic stars, it would be an O8V. The
corresponding luminosity and mass would be {\it log
L}\,=\,5.235\,$L_{\odot}$\, and $M$\,=\,30\,\sm.  The reddening
correction for star \#29 is more important since the star is embedded
in a region where the extinction reaches a value of $A_{V}$\,=\,2.5
mag. Correcting for the reddening the absolute magnitude of the star
is estimated as $M_{V}$\,\ab\,--5, which corresponds to a main
sequence O6 star with a ZAMS mass of 45 \sm (Vacca et
al. \cite{vacca}). \\

Apart from the main sequence, the C-M diagram shows several stellar
populations with a variety of red colors.  These should be a mix of
stars in different stages of evolution affected by the
interstellar dust along the line of sight.  Disentangling those
populations is not an easy task and doing so using just the present
data would be rather speculative.  \\

\begin{figure*}
\begin{center}
\resizebox{17cm}{!}{\includegraphics{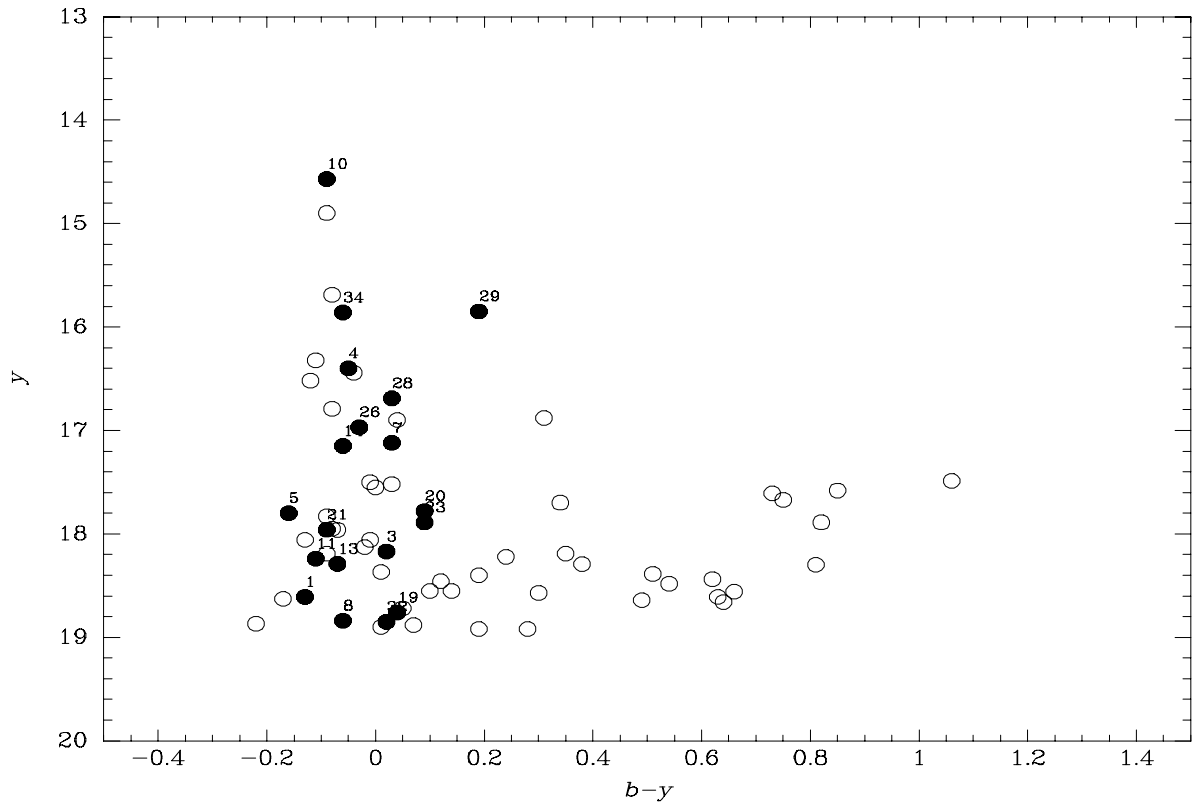}}
\caption{Color-magnitude diagram of the brightest stars 
(lower cutoff at $y=19$ mag) observed towards the \h2\, region N83B
based on WFPC2 imaging with the Str\"omgren filters $b$ (F467M) and
$y$ (F547M). The magnitudes are not corrected for the reddening.
A solid dot indicates a stars lying in the PC field (i.e. the area
surrounding N83B), while an open circle marks a star found in the
three WF CCDs.}
\label{cm}
\end{center}
\end{figure*}

\section{Discussion}

It is interesting to speculate about the physical conditions that led
to the formation of the two adjacent ionized blobs (N83B-1 and N83B-2)
inside the larger N83B and understand why they have different
sizes. Of course the size depends on the mass or temperature of the
exciting stars, but also on the density of the molecular clumps. On
the other hand, the mass of the exciting stars themselves might depend
on the clump density in which they formed.  So, was the initial
density of the molecular clumps the main factor, denser clumps formed
more massive stars, or were there random factors at work in a very
turbulent environment? We cannot provide a definite answer to all 
these questions, but we will try  to probe some of the issues
by placing these objects into the global star formation
history of the region.  To do so let us examine the N83 region in more
detail. \\

The N83 ionized gas complex occupies the central part of the LMC OB
association \lh5\, (Lucke \& Hodge \cite{lh}) and is dominated by the
bright and large \h2\, region N83A (Fig. 1).  The main exciting star
of N83A, Sk-69$^{\circ}$25 (Sanduleak \cite{sk}), is an O6 V((f)) type
with $V=12.28$, $B-V = -0.14$, and $U-B = -0.97$ mag (Massey et
al. \cite{massey}; Schmidt-Kaler et al. \cite{schmidt}). An age of
2.24 Myr and a ZAMS mass of 64 \sm\, are estimated for this star
(Massey et al. \cite{massey}).  Unfortunately little is known about
the stellar content and detailed physical properties of N83A.  We do
know however that it is associated with a large quantity of dust, as
shown by the IRAS observations or by its strong extinction (Caplan \&
Deharveng
\cite{cd}, Ye \cite{ye}). An \ha\,/\hb\, ratio of 3.77 derived by
these authors for the whole N83A indicates an interstellar extinction
of $A_{V}$\,=\,0.8 mag. \\

On the other hand, N83B which lies at \ab\,2\min\, (\ab\,30 pc)
north-east of N83A (Fig.\,1), probably represents the most recent
massive starburst in the giant \h2\, complex N83 and the OB
association \lh5. The young age is inferred from the fact that the
burst is associated with a compact high-excitation \h2\, blob, N83B-1,
which is strongly affected by local dust. The extreme youth of N83B is
corroborated by the results of Israel \& Koornneef (\cite{ik91}) who
conducted a {\it JHK} survey of the LMC \h2\, regions and compared it
with the IRAS observations. In the case of N83B, they detected a
compact source of less than 10\frac\, in size at both near-IR and IRAS
wavelengths, whereas no compact source was found towards N83A.  An
inspection of their Table 3 shows that all very young star formation
regions are associated with compact IR sources, as is particularly the
case for other Magellanic \h2\, blobs, such as N159-5, N81,
and N88A (Heydari-Malayeri et al. 1999 a,b,c), as well as for
many Galactic \h2\, regions. \\

The {\it HST} observations reveal those, so far hidden,
components of N83B and bring to light a starburst made of dozens of
newborn stars.  As discussed earlier, the star cluster
consists of a well-defined main sequence population with 19 members
ranging in magnitude from $y=14.57$ to 19.00 spreading over 56
pc$^{2}$.  They are all associated with the \h2\, region and many of
them participate in its ionization. Our observations also
imply that the true number of massive stars in the LMC is
underestimated, since a large number of them should be hidden in
similarly unresolved small clusters. Based on the
number of detected stars over the area imaged by HST we estimate a
surface density of 0.34 stars pc$^{-2}$ for massive stars in N83B,
which is quite high, a factor of \ab\,50 higher than that for
\lh5\, (Lucke \cite{lucke}), and even a factor of 17 higher than the
mean density derived for a sample of five OB associations in the
Magellanic Clouds (Massey et al. \cite{massey95}, Hunter
\cite{hunter}). Though the total number of massive stars is not
well determined, even if we reduce them to 10 the
corresponding density is 0.18 stars pc$^{-2}$ which is still \ab\,10
time larger than that of the Magellanic Clouds OB associations. The
star density in the N83B-1 area is in fact higher than that, since we
are quite certain that at least stars \#10, \#28, \#29, and \#34 are
all massive and hot occupying an area of only 7 pc$^{2}$.
This yields a density of 0.57 stars pc$^{-2}$ which is interestingly
just  \ab\,3 times smaller than that of R136 (Hunter
\cite{hunter}). One should keep in mind though that the above
comparisons may not reveal the whole picture as they may be affected
by the small-number statistics due to the size of our regions and also
by the uncertainties on the number of candidate massive stars. On the
other hand, a recent study by Parker et al. (2001) shows that field
stars significantly contaminate the massive star population of the LMC
OB associations, and hence the corresponding initial mass functions.
Consequently, this tends to aggravate the discrepancy in the
estimated stellar densities. \\

Furthermore, our imaging clearly shows the structure of the
two compact \h2\, blobs, N83B-1 and N83B-2, and the small arc-nebula,
N83B-3, bathing in N83B.  The first one represents the brightest and
the most excited region of N83B.  Its exciting star, \#29, is
presumably the most massive and hottest star of the burst, since it is
responsible for the particularly high excitation measured from the
strength of the doubly ionized oxygen line.  As for the smaller blob
N83B-2, its ionizing source (\#28) is expected to be a late type O or
an early B, since the nebular \oiii\,/\hb\, ratio is not large.  The
same should be true for \#34, the exciting star of N83B-3. The
relatively high brightness of the arc south of this nebula suggests
the presence of high density material also in that direction. \\

The blob N83B-1 is likely the most obscured part of the N83 complex
with $A_{V}$\,\ab\,2.5 mag. It is probably the optical counterpart of
the compact IR source detected by Israel \& Koornneeff
(\cite{ik91}). While dust may be mixed with ionized gas, such a large
extinction implies the presence of dust also outside the nebula, since
the Balmer decrement due to absorption by internal dust becomes
saturated at a ratio of 4.3 (for $R=A_{V}/E(B-V)=3.1$) or $A_{V}=1.1$
mag.  Note also that the \h2\, region's mean \ha\,/\hb\, ratio of 3.85
is higher than the value of 3.77 obtained for N83A (Caplan
\& Deharveng \cite{cd}, Ye \cite{ye}).  An extinction of
$A_{V}$\,=\,0.5 mag derived for Sk-69$^{\circ}$25 from its photometry
and spectral type indicates that the main exciting star of N83A is not
situated in the most reddened part of the \h2\, region, contrarily to
what is the case for \#29, the exciting star of N83B-1. \\

A rather noteworthy point is the presence of the three \h2\,
region N83B-1, 2, and 3 at the same side of N83B. This
alignment need not be by chance, but probably due to the
presence of the adjacent molecular cloud. The particularly
high extinction of N83B-1 can be explained by the presence of dust
associated with the molecular cloud very probably located east of
N83B-1 and south of N83B-3.  In fact the brightest area of the nebula,
the blob N83B-1, is expected to lie near the intersection zone between
the ionized gas and the molecular cloud.  Two different models can be
proposed to explain the observed morphology. According to the first
one, N83B-1 and the shell nebula may be two parts of the same \h2\,
region. In that case the \h2\, region is probably ionization-bounded
towards N83B-1, while ionized gas is pouring out into the interstellar
medium mainly via the diffuse component, as predicted by the champagne
model (Tenorio-Tagle \cite{teno}, Bodenheimer et
al. \cite{boden}). Alternatively, we may be watching four separate
\h2\, regions: N83B-1 which is still embedded in the molecular cloud,
N82B-2, N83B-3, and the shell nebula, which should be a relatively
evolved \h2\, region. \\

In the framework of the latter model, which we favor, N83B-1 is
expected to be younger than the diffuse shell nebula, because it is
compact, denser, and still embedded in the molecular cloud. This leads
to the conjecture that the exciting stars of N83B-1 and N83B-2,
i.e. \#29 and \#28, may have been triggered by the eastward expansion
of the shell nebula according to the sequential star formation model
(Elmegreen \& Lada \cite{el}). There are many cases in the literature
for which this mechanism has been put forward to explain the
observations, in particular that of the SMC \h2\, region N88A
(Heydari-Malayeri et al. \cite{hey1999b}). \\

A conspicuous feature of N83B uncovered by the {\it HST} images is the
cavity carved in the diffuse nebula by the strong wind of the central
star \#10 which has a particularly strong ultraviolet radiation
(Table\,\ref{phot}). We try here to estimate an age for the
cavity. Assuming a typical mass loss rate of 6\,\x\,10$^{-5}$
\sm\,\,yr$^{-1}$ for the presumed central O star, a conservative wind
velocity of 1000 km s$^{-1}$, a gas density of 100 cm$^{-3}$, and an
observed radius of 2.5 pc, we find a lifetime of 37,000 yr from the
classical equations governing the interaction of the stellar wind and
the interstellar medium (Weaver et al. \cite{weaver}, Dyson
\cite{dyson}). Should the wind velocity be higher, 2000 km s$^{-1}$,
the corresponding age will be 23,000 yr. In the same way, we can
calculate the age of the sharp emission ridge with the
assumption that it is created by star\#29. Using the same mass loss
rate as the one adopted for \#10, a wind velocity of 1000 km s$^{-1}$,
a gas density of 1000 cm$^{-3}$ (Paper I), and a projected radius of
0.6 pc, we derive an age of 7400 yr for the bright ridge. It is also
possible that the ridge in fact results from the interaction of the
winds of both stars  \#29 and \#10. Consequently, its true age may be
larger if the wind of star \#10 significantly decelerates the westward
advance of the ridge and/or if its assumed radius is actually
underestimated due to projection effects. \\

These new {\it HST} data combined with large-scale,
low-resolution observations obtained with ground-based telescopes
provide new evidence in favor of hierarchical star
formation (Elmegreen \& Efremov \cite{elm96}, Elmegreen \cite{elm00})
in this region. According to this model, large-scale star-forming
regions have a fractal architecture, in the sense that the cloud
structures appear self-similar over a wide range of scales.  The gas
structures range from superclouds (\ab\,10$^{7}$\,\sm) in which star
complexes form, to giant molecular clouds (10$^{3}$ to 10$^{6}$\,\sm)
in which clusters and OB associations form, to small molecular cores
($<$ 10$^{3}$\,\sm) in which individual and multiple stars form
(Elmegreen \& Efremov \cite{elm96}).  A recent study by Elmegreen
(\cite{elm00}) of the age difference versus size has yielded an
interesting correlation for 244 LMC clusters with size and age ranges
\ab\,0.1--1000 pc and 10--100 Myr, respectively. It turns out
that the timescale for coherent star formation in a region is always
about one turbulent crossing time (defined as half-size divided by the
Gaussian dispersion in internal velocity), scaling approximately with
square root of size.  This implies that small regions should evolve
faster than large regions: inside the larger regions, numerous smaller
regions form and dissolve many times over a relatively long
period. \\

This scenario seems very attractive for the OB association \lh5\, and
the \h2\, complex N83. In other words the {\it HST} observations
provide a rare case in which by penetrating into very small spatial
scales in the LMC we discover a fractal star-forming region.  The OB
association \lh5, which spreads over an area of 16 square minutes
(3600 square pc), harbors tens of bright, blue and red stars as well
as several \h2\, regions.  None of the brightest stars lie in the {\it
HST} field of view, except the ones indicated in Fig. 1. The brightest
star of the association, Sk-69$^{\circ}$30 (Sanduleak \cite{sk}), is
in fact an evolved supergiant of type G5 Ia (Massey et
al. \cite{massey}) with $V=10.18$ mag (Schmidt-Kaler et
al. \cite{schmidt}).  The association contains also a Wolf-Rayet star
of type WN2, called BAT99-5 (Conti \& Massey \cite{c/m}; Breysacher et
al. \cite{bat}), which naturally comes from the evolution of a massive
O type progenitor.  Using photometry and spectroscopy to construct the
HR diagram and by fitting isochrones, Massey et al. (\cite{massey})
estimate an age of 1.26 to 2.24 Myr for the brightest main sequence O
type stars of the association, while they derive an age of 6.46 Myr
for the supergiant Sk-69$^{\circ}$30 (Fig.\,\ref{lh5}).  On the other
hand, almost all of the known O type stars of \lh5\, either lie
outside the \h2\, regions, or are associated with diffuse, low-density
ionized gas, implying that these stars have had enough time to fully
or partially dissipate their \h2\, regions.  All these facts
support a scenario of continuous massive star formation with
an age spread in the OB association \lh5, in agreement with a
timescale of \ab\,9 Myr estimated from Elmegreen's (\cite{elm00})
hierarchical scaling law. \\

At a lower level, N83A, the largest \h2\, region of the association,
extends over \ab\,2\min\, or \ab\,30 pc. Let us assume that the \h2\,
region traces the size of the molecular clump which formed
the stars in N83A. One may argue that this is not justified if the
associated burst is tightly packed and the hot gas expands
fast. However, the diffuse structure of N83A which shows no
conspicuous brightness peak supports the assumption. Using the same
law, a timescale of \ab\,6 Myr is therefore expected for the star
formation in N83A, and this is in agreement with the above-mentioned
age estimate of 2.24 Myr for Sk-69$^{\circ}$25, the main exciting star
of N83A. \\

The brighter and more compact \h2\, region N83B has a diameter of
\ab\,30\frac\, or 7.5 pc, and the uncovered main sequence blue stars
extend over almost the same area. Therefore the
hierarchical scaling law predicts a younger age of \ab\,3 Myr for this
region which is consistent with the evidence we presented in this
paper suggesting that N83B is in fact younger than N83A. \\

Going down towards even smaller scales, the presence of the two blobs
N83B-1 and N83B-2 and the arc-nebula inside N83B advocates the idea of
a fractal structure in the molecular cloud which gave rise to star
formation in this part of the LMC. This obviously presumes that even
the sequential star formation needs a pre-existing gas concentration,
since it only provokes the collapse of the gas and will be inoperative
in its absence.  The corresponding timescale for N83B-1, as
predicted by the hierarchical model, is \ab\,1 Myr and follows our
scenario in which we have argued that N83B-1 is the youngest
and most massive product of star formation in this region. \\

If the described size-age relation is valid at even lower spatial
scales, the \h2\, region N83B-2 should be younger than N83B-1, as its
size is smaller. However, at this stage we have no solid evidence on
the age of N83B-2 which would corroborate this assertion. As discussed
earlier, the \oiii\,/\hb\, ratios indicate that the exciting star of 
N83B-2 (\#28) may actually be less massive than the one inside N83B-1
(\#29). Should  this be the case, this does not contradict the scenario
of fractal star forming structures, as it is possible that in that
location the physical conditions (e.g. lack of a large molecular
cloud) were not favorable for the formation of a more massive star.
Some of these uncertainties may be resolved in the future if spectra of
these stars are obtained and accurate ages are estimated.

\section{Conclusions}

Based on the new {\it HST} observations of N83B described in the 
paper, we conclude the following: \\

1) The so-called high excitation blob N83B is resolved into two extremely
compact regions N83B-1 and N83B-2. Together with the arc-nebula
N83B-3, all three are located in the eastern edge of N83B and
their origin appear to be the result of triggered/sequential star
formation. \\

2) A large number of candidate massive stars is revealed
making the stellar surface density of the region several times 
larger than what is typically found in other the OB associations of the
Magellanic Clouds. \\

3) Taking into account the observations ranging from large scales (of
order 100 pc, see Fig.\,1) to small scales (of order 1 pc, Fig.\,3)
seems to indicate that indeed the age of a star formation region
scales with the square root of its size (1 Myr for 1 pc). The
underlying reason for this scaling relation (``star formation law'')
could be sought in the fact that the mean gas density of a star
forming region is inversely proportional to its linear dimension such
that the mean gas column density is constant for all scales. Since the
dynamical timescale is inversely proportional to the square root of
the mean density (1 Myr for 10$^{3}$ cm$^{-3}$), the dynamical time
would scale with the square root of the size. \\

\begin{acknowledgements} 
The authors are grateful to the referee Dr. Joel Wm. Parker
(Southwest Research Institute) whose useful comments contributed to
improve the paper. We are also indebted to Dr. Bruce Elmegreen 
(IBM Research Division) for 
reading the manuscript and discussion.
VC would like to acknowledge the financial support
for this work provided by NASA through grant number GO-8247 from the
STScI, which is operated by the Association of Universities for
Research in Astronomy, Inc., under NASA contract 26555.
\end{acknowledgements}

{}

\end{document}